\documentclass[aps,floats,twocolumn,preprintnumbers]{revtex4}
\usepackage{epsfig}
\usepackage{amsmath}
\usepackage{amssymb}
\usepackage{amsthm}
\usepackage{graphics}
\usepackage{float}
\usepackage{color}
\usepackage{graphicx}

\newcommand{\bea}{\begin{eqnarray}}
\newcommand{\eea}{\end{eqnarray}}
\newcommand{\beq}{\begin{equation}}
\newcommand{\eeq}{\end{equation}}

\DeclareMathAlphabet{\mathpzc}{OT1}{pzc}{m}{it}

\begin{document}
\title{Spin polarization effects in micro black hole evaporation}

\author{Antonino Flachi}
\email{flachi_AT_yukawa.kyoto-u.ac.jp} 
\author{Misao Sasaki}
\email{misao_AT_yukawa.kyoto-u.ac.jp} 
\author{Takahiro Tanaka}
\email{tanaka_AT_yukawa.kyoto-u.ac.jp} 
\affiliation{Yukawa Institute for Theoretical Physics, Kyoto University, Kyoto 606-8502, Japan}
\preprint{YITP-08-71}

\begin{abstract}
We consider the evaporation of rotating micro black holes 
produced in highly energetic particle collisions, 
taking into account the polarization due to
 the coupling between the spin of the emitted particles and the angular
 momentum of the black hole. 
The effect of rotation shows up in the helicity dependent angular distribution 
significantly. By using this effect, there is a possibility to determine the axis 
of rotation for each black hole formed, suggesting a way 
to improve the statistics. 
Deviation from thermal spectrum is also a signature of rotation. 
This deviation is due to the fact that 
rapidly rotating holes have 
an effective temperature $T_{\rm eff}$ significantly higher 
than the Hawking temperature $T_H$. 
The deformation of the spectral shape becomes 
evident only for very rapidly rotating cases. 
We show that, since the spectrum follows a blackbody profile with an 
effective temperature, it is difficult to determine both 
the number of extra-dimensions and the rotation parameter 
from the energy spectrum alone. 
We argue that the helicity dependent angular distribution 
may provide a way to resolve this degeneracy. 
We illustrate the above results for the case of fermions. 
\end{abstract}
\maketitle

\vspace{2mm}
{\it Introduction}. 
Within the context of TeV-scale
gravity~\cite{add,aadd,rs}, the possibility that colliders or cosmic ray 
facilities may observe micro black holes has attracted enormous 
attention~\cite{dl,gt,arg,fs,kalter}. A close look at the limits on the
fundamental Planck scale shows that
a window of about $5$ TeV is still open for the LHC to observe such 
exotic events~\cite{kantilast}, while
the window is much wider for cosmic rays. 
Micro black holes with even higher energies could be produced from the collision of
a cosmic ray with an atmospheric nucleon, a dark matter particle, or
another cosmic ray (Ref.~\cite{masip} gives some up-to-date estimates). 

In this paper, we consider micro black holes resulting from the
collision of two particles at energies much higher 
than the higher dimensional Planck mass $M_P$. We have in mind
models with $M_P$ of order of a few TeV and the standard
model confined on a $3$-brane, embedded in a $(4+n)$-dimensional bulk. 
These black holes have 
horizon radius smaller than the size of the extra dimensions, and are
expected to follow balding, spin-down, Schwarzschild, and Planck
phases. 
Micro black hole formation has been studied both analytically~\cite{eg}
and numerically~\cite{yn,yr}, 
and their evaporation has also been the subject of considerable attention 
(See Refs.~\cite{kantilast,kanti_review} for review). 
Previous work suggests that micro black holes will emit mostly brane
modes~\cite{ehm,ccg}, and 
the deviations from the blackbody spectrum 
have been investigated
 using numerical and semi-analytical methods~\cite{kerr_emission_0,kerr_emission_1,kerr_emission_2,kerr_emission_3,kerr_emission_4,kerr_emission_5,kerr_emission_6,kerr_emission_7}. 

\vspace{1mm}
{\it Radiation.} In this paper we analyze the fermion 
emission from spinning evaporating micro black holes. Assuming that the
black hole 
horizon is significantly smaller than the extra dimensions, 
we approximate it by a vacuum higher dimensional Kerr~\cite{mp}:
\begin{eqnarray}
ds^2 &=& \left(1-{M\over \Sigma r^{n-1}} \right) dt^2
 + {2aM \sin^2 \theta \over \Sigma r^{n-1}} dt d\varphi - {\Sigma\over \Delta}dr^2
\cr
&&-\Sigma d\theta^2  -\left(r^2+a^2+{a^2M \sin^2\theta\over \Sigma r^{n-1}} \right) 
\sin^2\theta d\varphi^2 \cr
&&-r^2 \cos^2\theta d\Omega_n^2~,\nonumber
\end{eqnarray}
where $\Delta \equiv r^2+a^2 - M r^{1-n}$ and $\Sigma \equiv r^2+a^2\cos^2\theta$. $M_P$ is normalized to one. 
Since we are interested in the visible brane modes, the
background spacetime will be given by the projection of the above metric 
on the brane. 

Massless fermions emitted by the black hole are described by the Dirac equation:
$$
e_a^\mu \gamma^{a}\left(\partial_\mu +\Gamma_\mu\right)\psi=0~,
$$
where $\psi$ is the Dirac spinor wave function, $e_a^\mu$ a set of tetrads, 
$\Gamma_\mu$ the spin-affine connections determined by $\Gamma_\mu = \gamma^a\gamma^b\omega_{ab\mu}/4$~, with $\omega_{ab\mu}$ being the Ricci rotation coefficients. The matrices 
$\gamma^{\mu}=e_a^\mu \gamma^{a}$ are chosen to satisfy the relation
$\gamma^\mu\gamma^\nu+\gamma^\nu\gamma^\mu= g^{\mu\nu}$, with
$g^{\mu\nu}$ being the metric on the brane.

The Dirac equation for massless fermions on a Kerr background has been studied
extensively in four and higher 
dimensions~\cite{vi,unruh1,pt,page,kerr_emission_1,kerr_emission_6,kerr_emission_7}. 
Here we briefly describe the calculation, closely following the approach
of Ref.~\cite{unruh1}. Due to the symmetries of the Kerr spacetime, 
the spinor wave function factorizes as
\beq
\psi = {\mathpzc N}{e^{i(m\varphi-\omega t)}}
\left( \begin{array}{c}
\vec{\phi} \\
\pm \vec{\phi} \end{array} \right)~,
\nonumber
\eeq
where the $+$ and $-$ signs refer to negative and positive helicities, respectively.
We illustrate the results for the case of negative helicity.
The positive helicity case can be obtained by a trivial chirality
transformation. 
The field $\vec{\phi}$ takes the form
\beq
\vec{\phi}= 
\left( \begin{array}{c}
{ R}_{-}(r){S}_{-}(\theta)\\
{R}_{+}(r){S}_{+}(\theta) \end{array} \right)~,
\nonumber
\eeq
and the normalization factor is
$
{\mathpzc N}^{-1} =  \Delta^{1/4}(r+i a \cos\theta)^{1/2} \sin^{1/2}\theta.
$
The angular and radial modes obey
\begin{eqnarray}
\left({d\over d\theta} \pm\omega a \sin \theta \mp{m\over \sin\theta}\right)S_{\mp}(\theta)
&=&\pm\kappa S_{\pm}(\theta)~,
\nonumber\\
\left({d\over dr}\mp{i\over \Delta}\left( \omega (r^2+a^2)
 -ma\right)\right)R_{\mp}(r)&=&\kappa \Delta^{-1/2}R_{\pm}(r)~,
\nonumber
\end{eqnarray}
where $\kappa$ is a separation constant. Supplemented with
regularity conditions at $\theta=0$ and $\pi$, the set of angular equations
provides an eigenvalue problem, which determines $\kappa$~\cite{pt}. 
In order to compute the particle flux, we need the solution to the radial
equation supplemented by ingoing boundary conditions at the
horizon
$$
R_- \sim 0~,~~R_+ \sim e^{-i\tilde \omega r_*}~,~\mbox{for $r_*\rightarrow -\infty$}
$$
where $\tilde\omega=\omega -ma/(r_h^2+a^2)$, and $r_*$ is defined 
by $dr_*/dr = (r^2+a^2)/\Delta$.
The number of particles emitted, for fixed frequency $\omega$, is distributed
according to the Hawking radiation formula. For negative helicity modes,
the angular distribution reads:
\beq
{d N\over d\omega d\cos \theta} = 
{1 \over 2\pi\sin\theta} \sum_{l,m} |S_{-}(\theta)|^2
{\sigma_{l,m}\over e^{\tilde\omega/T_H}+1}~,
\label{flux}
\eeq
where
$
T_H={1\over 4\pi r_h}{(n+1)r_h^2+(n-1)a^2 \over r_h^2+a^2}
$
is the Hawking temperature, and the grey-body factor $\sigma^{}_{l,m}$ is 
the squared
amplitude of the transmission coefficient of an incoming wave from
$r=\infty$ (see Ref.~\cite{page}).
\begin{figure}
\hspace*{-0.2cm}
\scalebox{1}{\resizebox{4.25cm}{3.4cm}{\includegraphics{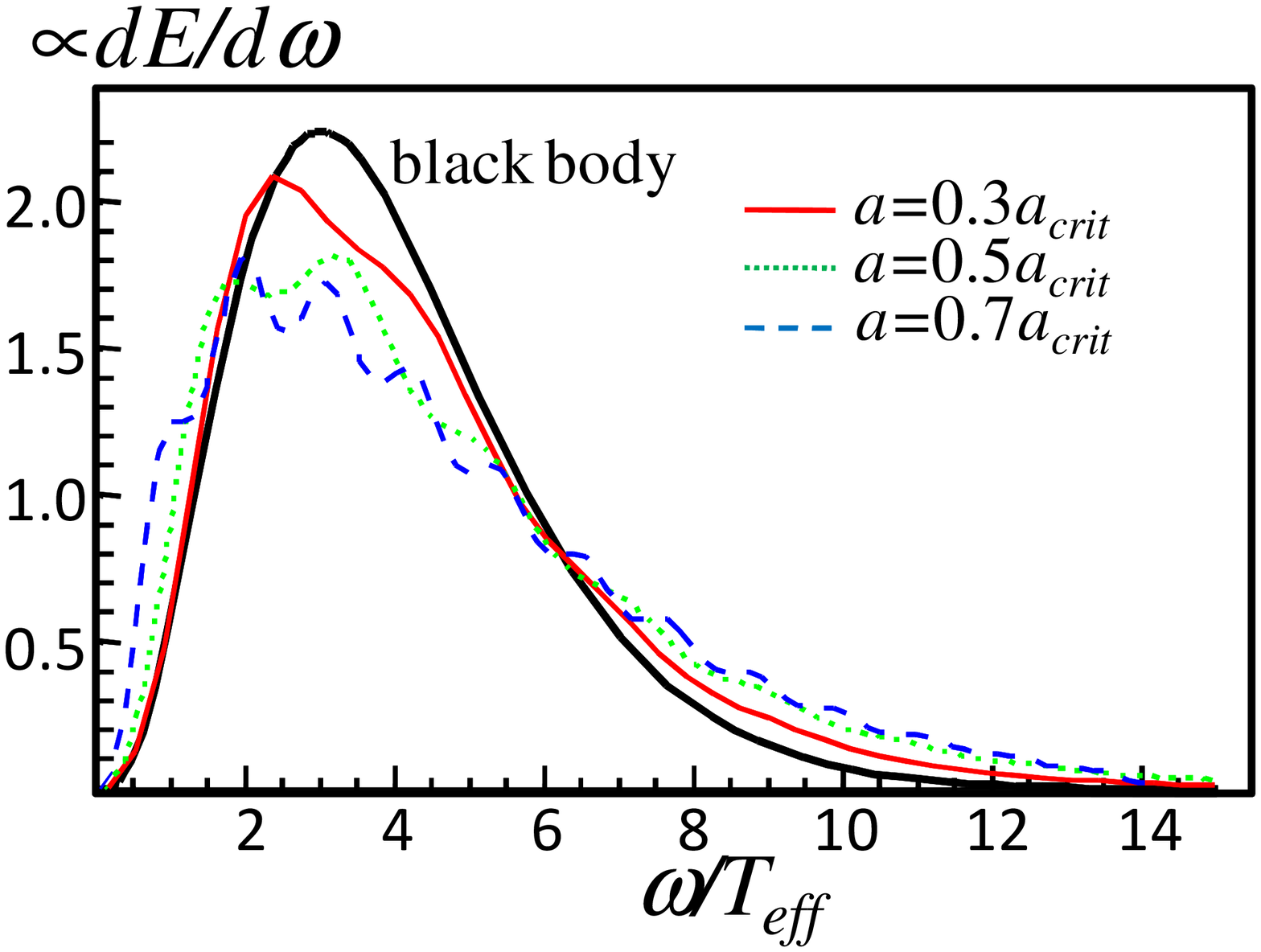}}}
\scalebox{1}{\resizebox{4.25cm}{3.4cm}{\includegraphics{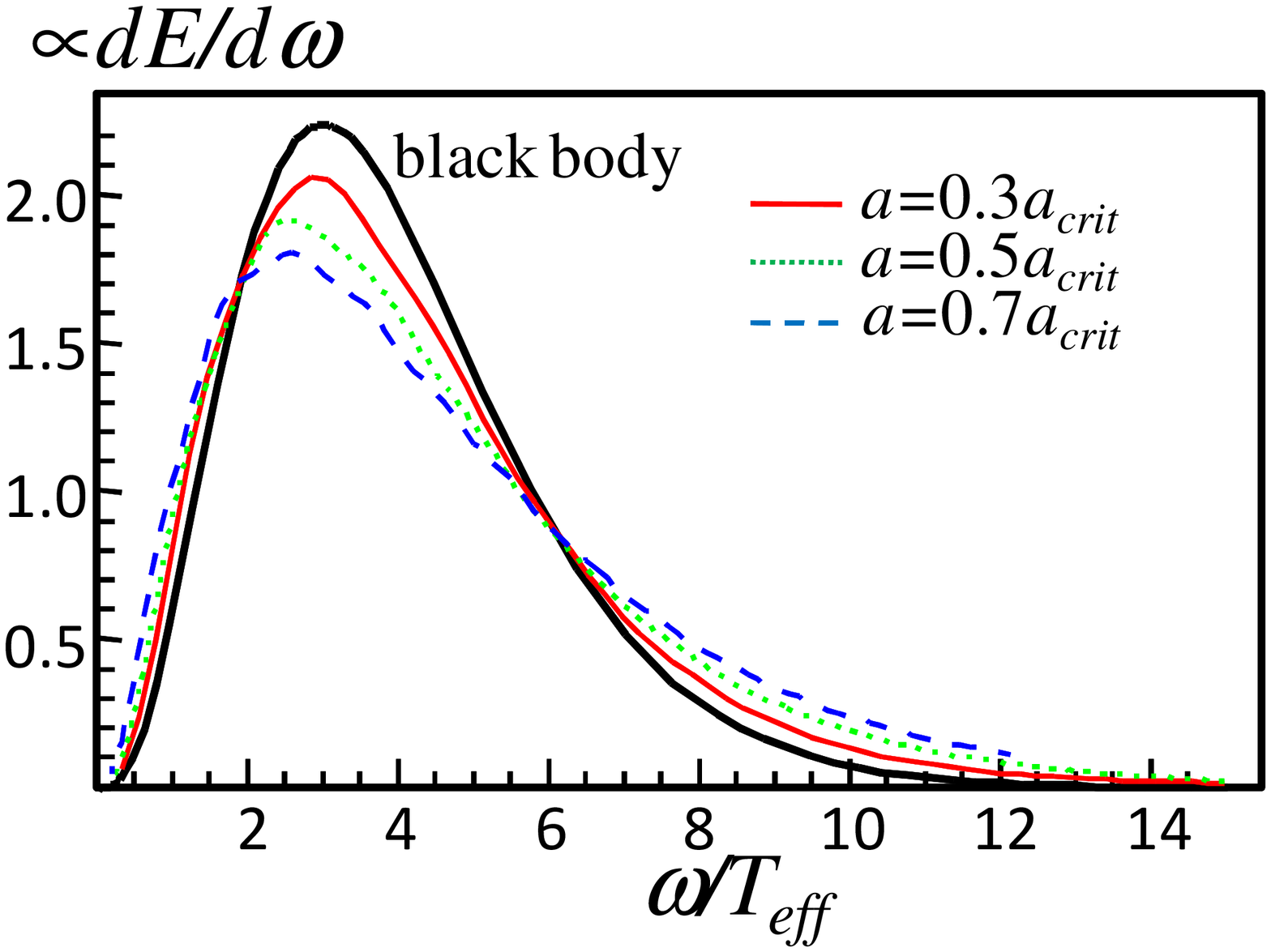}}}
\hspace*{-0.2cm}
\scalebox{1}{\resizebox{4.25cm}{3.4cm}{\includegraphics{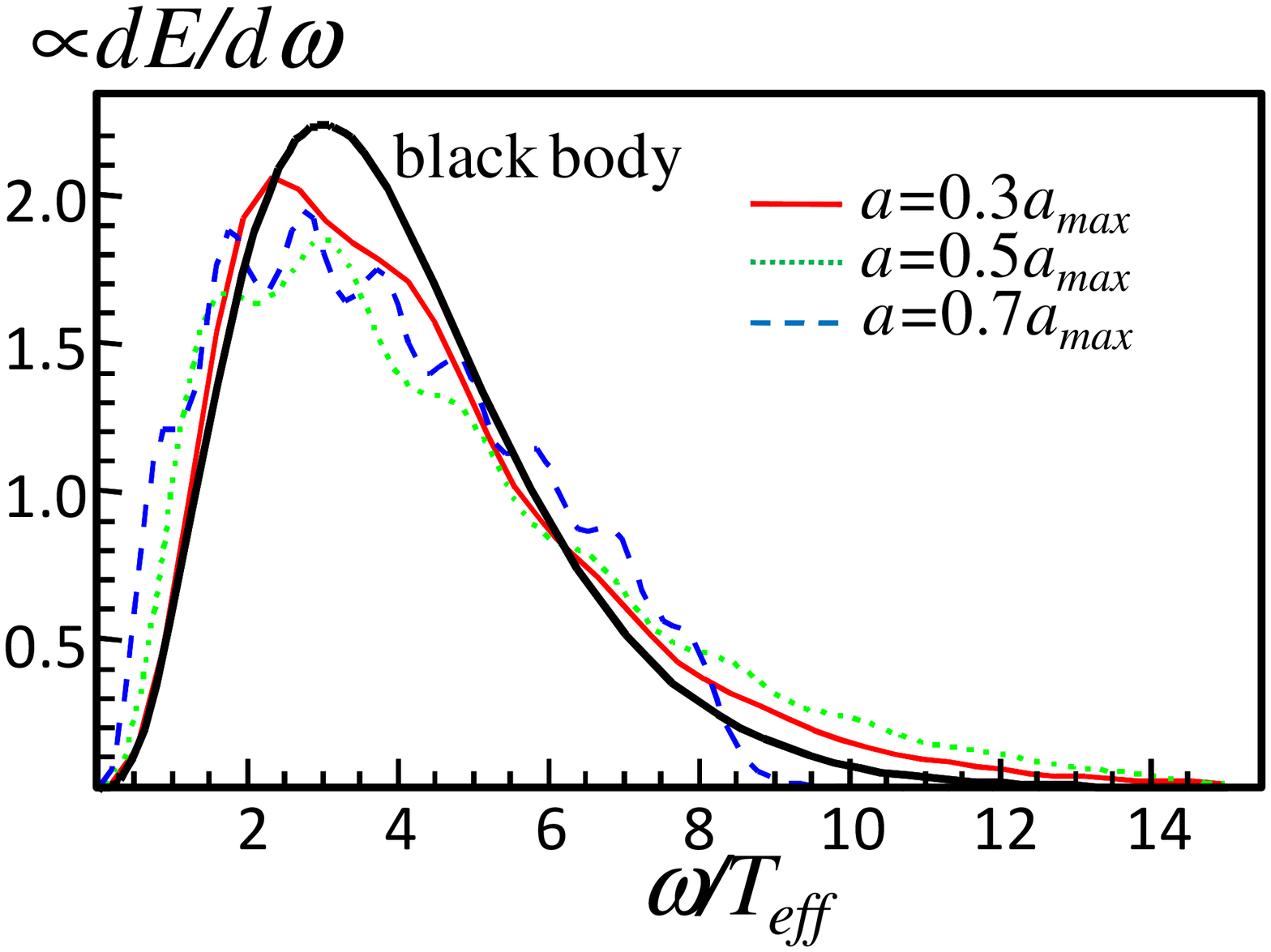}}}
\scalebox{1}{\resizebox{4.25cm}{3.4cm}{\includegraphics{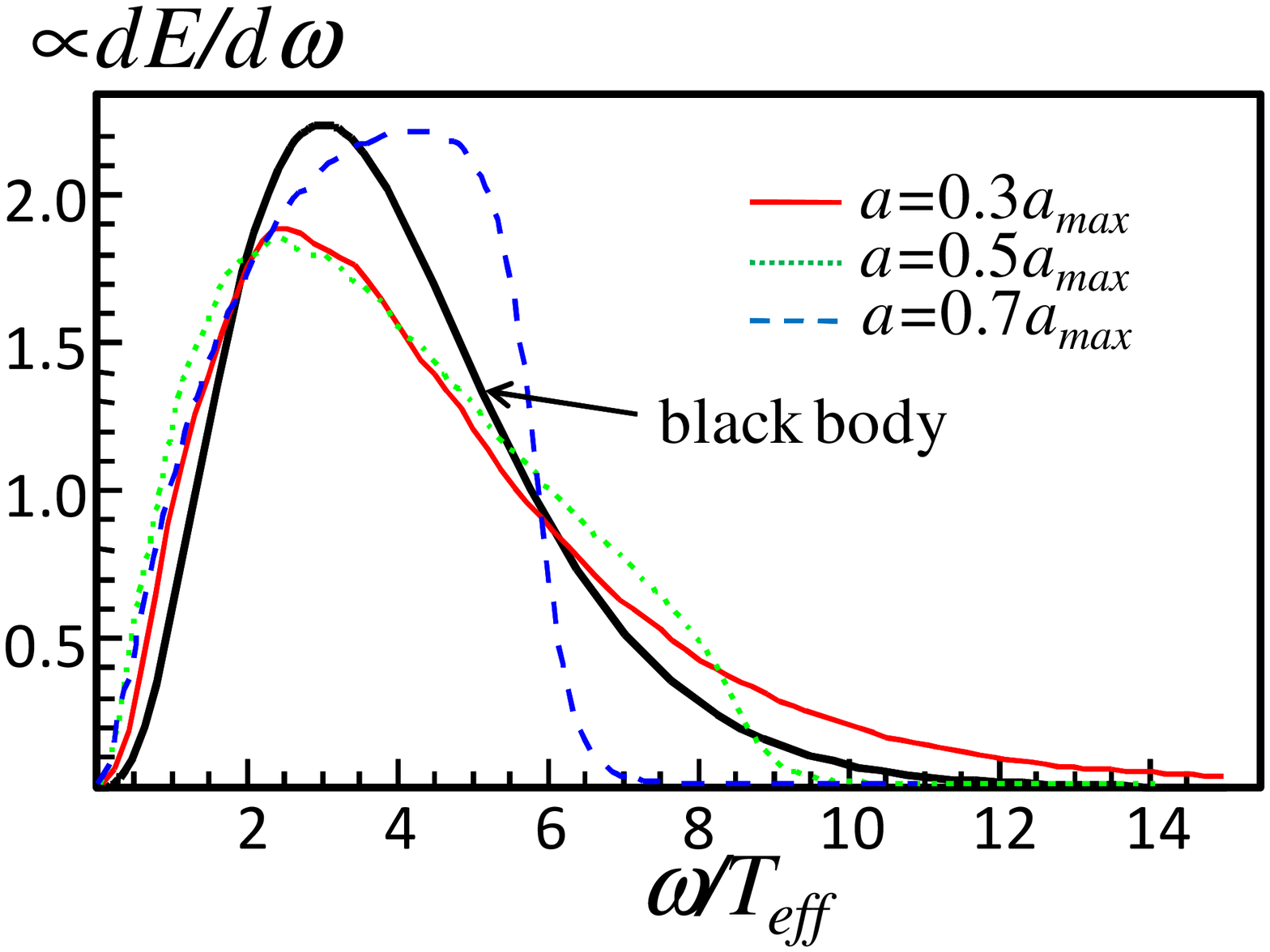}}}
\caption{
The normalized energy spectrum of the emitted fermions. 
The horizontal axis is rescaled by the effective temperature determined
by fitting the data by a black body profile. The overall amplitude
 is also normalized since the absolute magnitude is not observable. 
The upper and the lower panels are the plots for $n=2$ and $n=4$, respectively. 
The rotation parameter $a$ is set to $30\%,~50\%$ and $70\%$ of 
$a_{crit}$ (left) and  $a_{max}$ (right).}
\label{spectrum}
\end{figure}

\vspace{1mm}
{\it Results.} 
The initial angular momentum of the produced black holes $J=2aM/(n+2)$ is restricted by
requiring the impact parameter $b=J/M$ to be smaller than the horizon
radius $r_h$, determined by $\Delta(r_h)=0$. 
Then, the maximum value of the rotation parameter $a$ turns
out to be $a_{max}={n+2\over 2}r_h$~\cite{kerr_emission_1}. 
The upper bound on $J$ might be even lower 
for $n\geq 2$. 
In fact, there exists a critical value for $a$,
$a_{crit} \equiv (n+1)(n-1)^{-1}r_h^2$, 
where $|\partial(T,\Omega_h)/\partial(M,J)|$ vanishes. 
If the same argument as in the case of black branes applies, 
black holes with $a>a_{crit}$ suffer from the Gregory Laflamme
instability (See also Ref.~\cite{emparan}). 
Then, $a_{crit}$ represents the maximal value below
which the higher dimensional Kerr solution
is adequate. 
The value at which the dynamical instability is expected to set in 
may be slightly different from $a_{crit}$, which only represents an indicative estimate 
of the critical value.
Interestingly $a_{crit}<a_{max}$ (for $n=2,~3,~4$ extra
dimensions, $a_{crit}=1.09,~1.07,~1.06$, whereas
$a_{max}=1.25,~1.89,~2.46$). 
Although it is widely believed that a dynamical instability exists, 
the value of $a_{crit}$ obtained above is only heuristic. 
Thus, we consider two possible
cases: the maximal value allowed for $a$ is
$a_{crit}$ or $a_{max}$. 
A set of representative values for the
parameter $a$ is chosen as $a/a_{max}=~0.3$, $0.5$, $0.7$,
and $a/a_{crit}=~0.3$, $0.5$, $0.7$. 
$M$ is set to unity. 
Another natural choice to present the results would be to use the impact parameter $b$. In our case, we use the ratios $a/a_
{crit}$ and $a/a_{max}$, which correspond respectively to $b/b_{crit}$ and $b/b_{max}$, where $b_{crit}$ and $b_{max}$ are the values for the impact paramter corresponding to the critical and
 maximal cases respectively.

\begin{figure}
\hspace{-0.5cm}
\scalebox{1}{\resizebox{4.25cm}{3.4cm}{\includegraphics{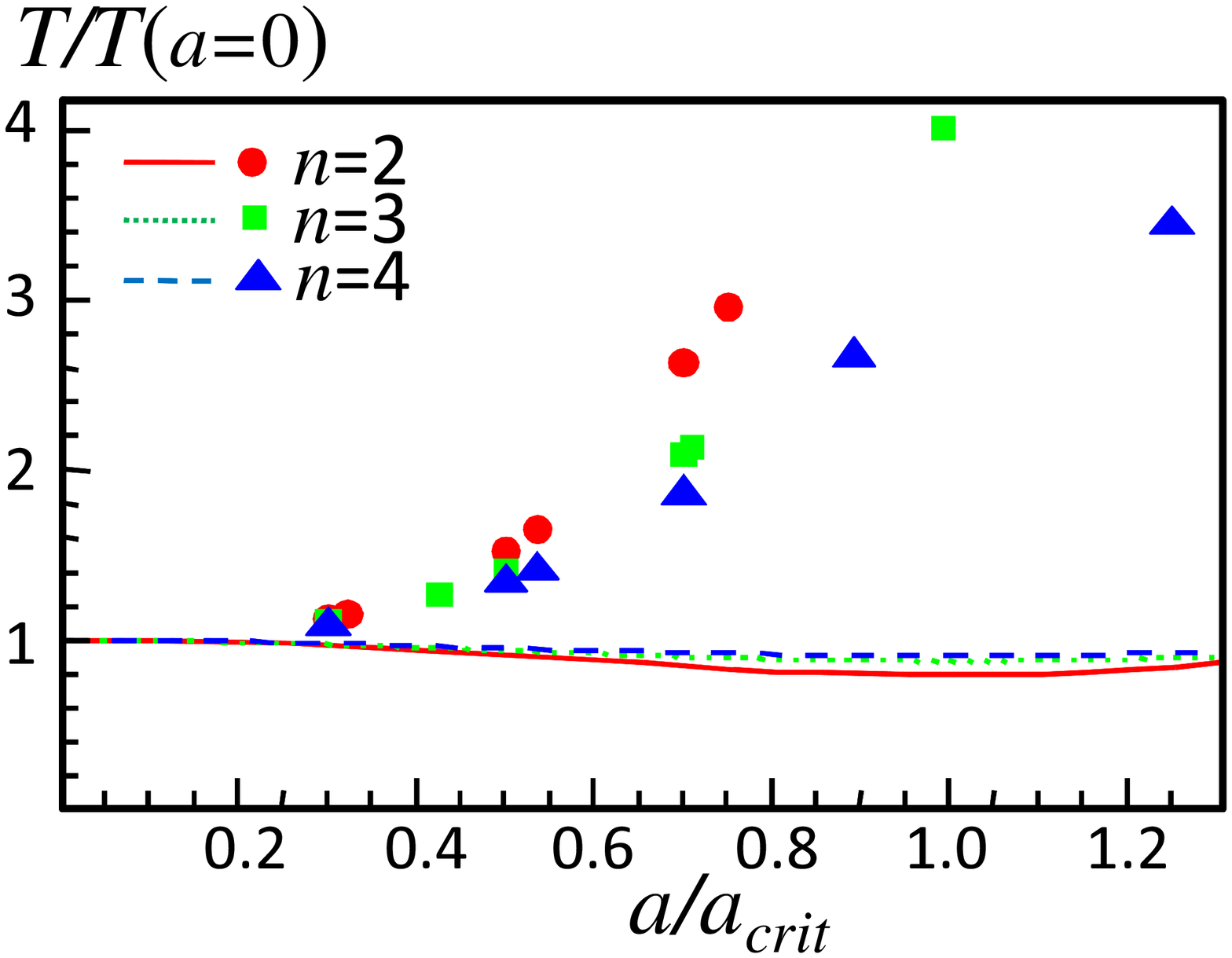}}}
\hspace{-0.1cm}
\scalebox{1}{\resizebox{4.25cm}{3.4cm}{\includegraphics{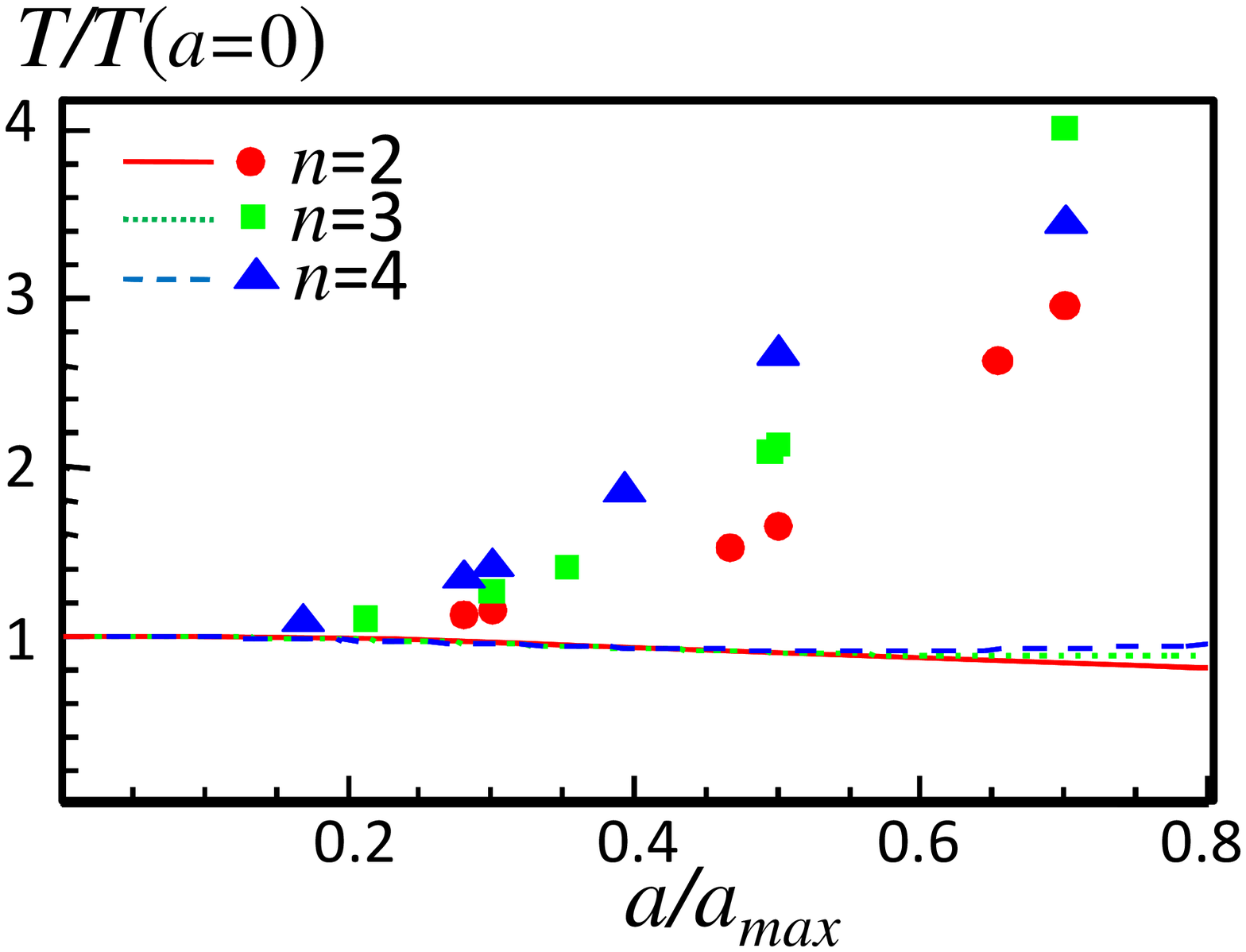}}}
\caption{The effective temperature normalized by the 
temperature at $a=0$ versus the rotation 
parameter $a$ normalized by $a_{crit}$ (left) and $a_{max}$ (right). 
The curves represent the Hawking temperature. }
\label{empiricalfit}
\end{figure}

Having fixed $a$ in the above way,
we compute the energy spectrum, shown in 
Fig.~\ref{spectrum}.  
We normalize the horizontal axis by using an effective temperature 
$T_{\rm eff}$ determined by fitting the data by a blackbody 
spectrum profile. 
The effective temperature 
$T_{\rm eff}$ is 
much higher than the 
Hawking temperature as shown in Fig.~\ref{empiricalfit}. 
However, 
the spectral shape is not so different from the thermal one
except for the cases with $a\approx a_{max}$ (Fig.~\ref{spectrum}).  
In previous work, \cite{kerr_emission_7,kerr_emission_6}, the enhancement of emission at large frequencies is reported. However, the deviation from a blackbody spectrum was not quantified.

We find that 
the renormalized spectra are enhanced for both lower and 
higher frequencies compared with the black body spectrum at $T=T_{\rm
eff}$ (thick line). 
Except for very large values of $a$, we have shown  that the obtained spectra can be fit well by superpositions of black body profiles with width of about $2\Omega_H\times T_{\rm eff}$. 
The intuitive reason for the enhancement of the effective temperature is that the motion of the hypothetical emitting 
 surface on the rotating black hole, relative to observers
 at infinity, causes an additional blueshift factor (which varies from place to place). This is because corotating emitted particles encounter less suppression from the statistical factor. This can be made precise by closely inspecting the combined behavior of the greybody and statistical factors. 
The dominant contribution to the spectrum comes from the $l=m$ modes and for larger values of the rotation the contribution to the spectrum from such modes, with large $l$, is non negligible. 

However, because of the 
change in the temperature and the rotation parameter 
during the evaporation, 
the broadening of the spectrum due to the rotation 
will not be identified straightforwardly.  
Wiggles can be also seen in the spectrum 
for a small number of extra-dimensions. However, 
wiggles are likely to disappear as $T$ and $a$ change during the 
evaporation. 

When the rotation velocity is high, the deviation from the 
thermal spectrum is much clearer. As a novel signature, 
we find that the spectrum is sharply cut off at high-frequencies for 
rapid rotation. 
This new signature may survive 
even after we take into account the superposition of spectra along the 
evolutionary track of an evaporating micro black hole. 
This highly spinning regime 
is realized for $a > a_{crit}$.  

\begin{figure*}
\scalebox{1}{\resizebox{4.75cm}{4.75cm}{\includegraphics{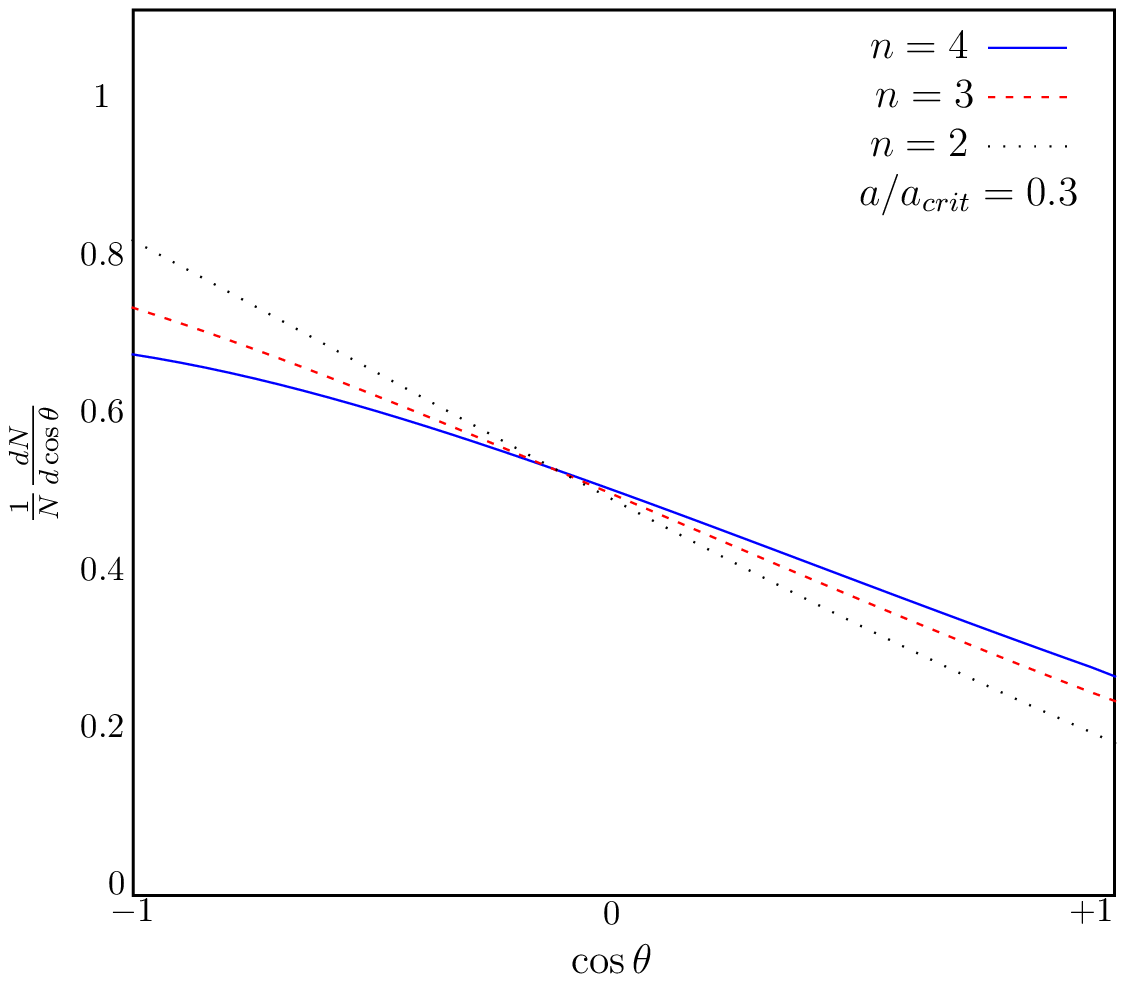}}}
\hspace{0.1cm}
\vspace{0.1cm}
\scalebox{1}{\resizebox{4.75cm}{4.75cm}{\includegraphics{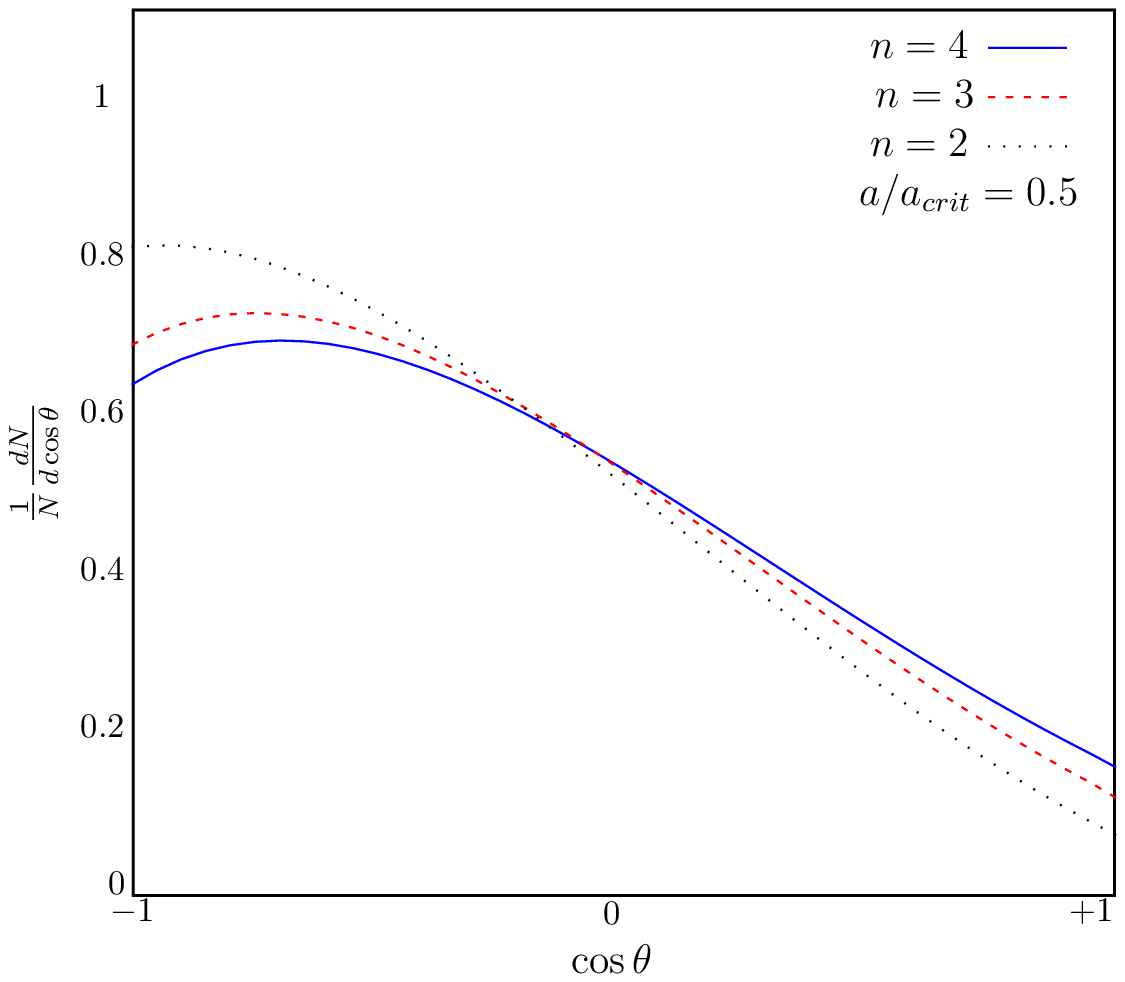}}}
\hspace{0.1cm}
\scalebox{1}{\resizebox{4.75cm}{4.75cm}{\includegraphics{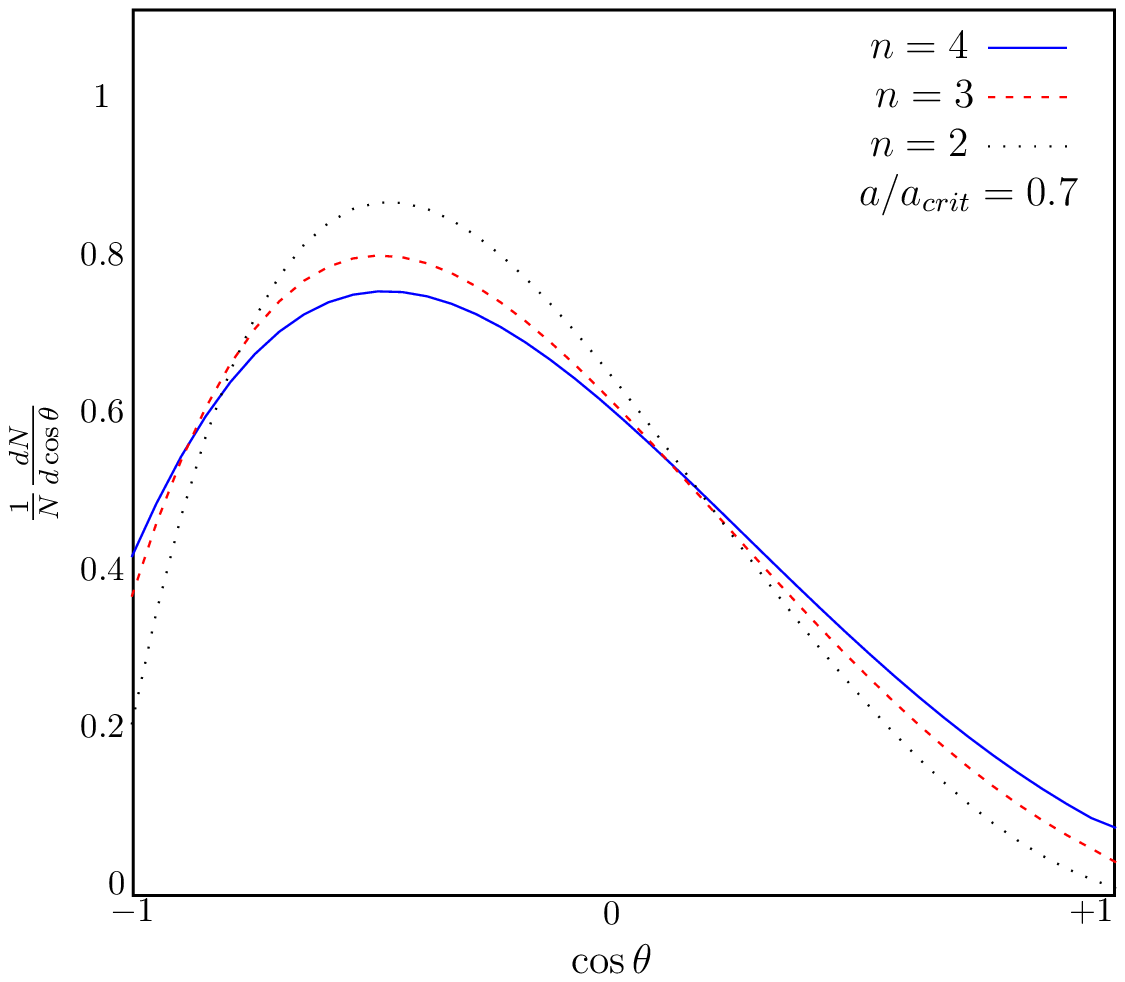}}}
\vspace{0.1cm}
\scalebox{1}{\resizebox{4.75cm}{4.75cm}{\includegraphics{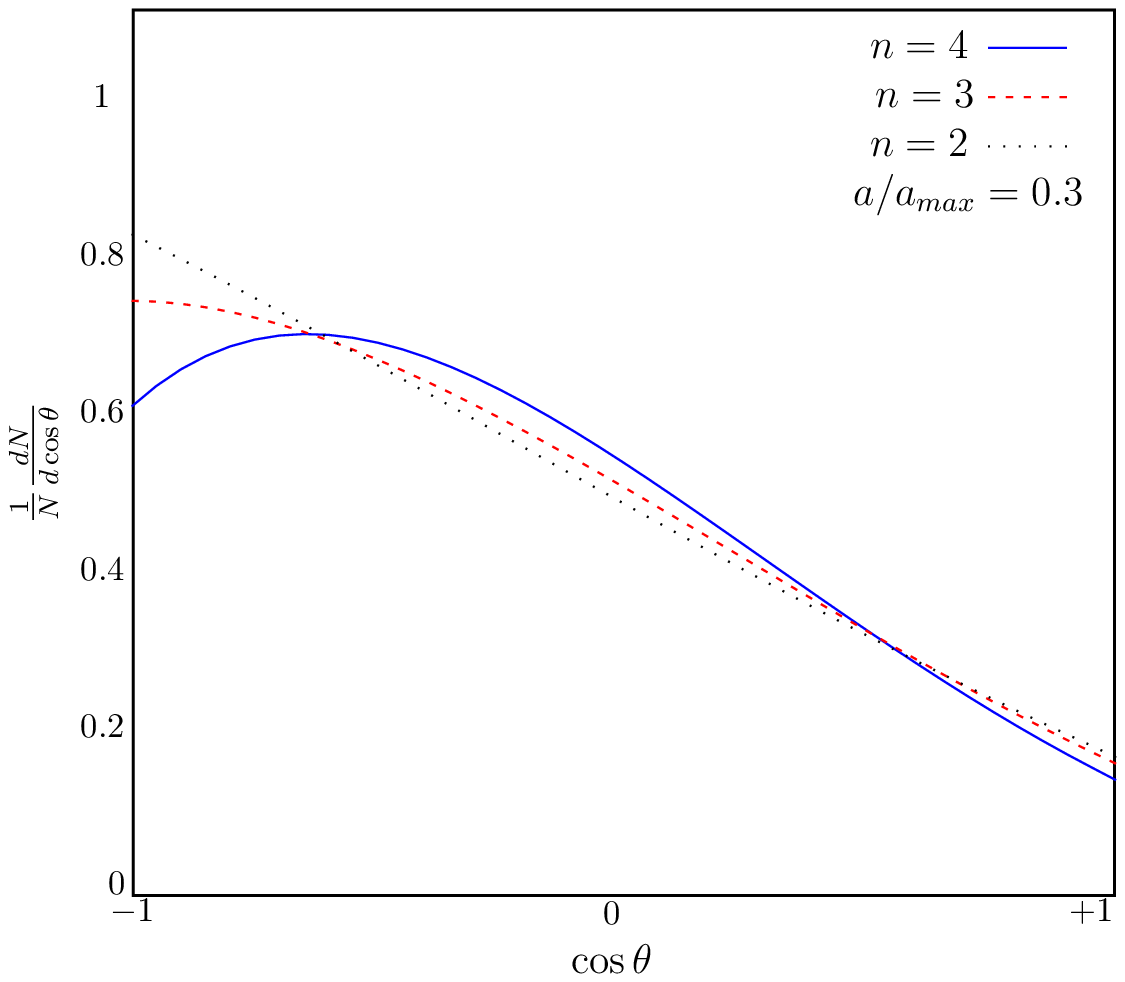}}}
\hspace{0.1cm}
\scalebox{1}{\resizebox{4.75cm}{4.78cm}{\includegraphics{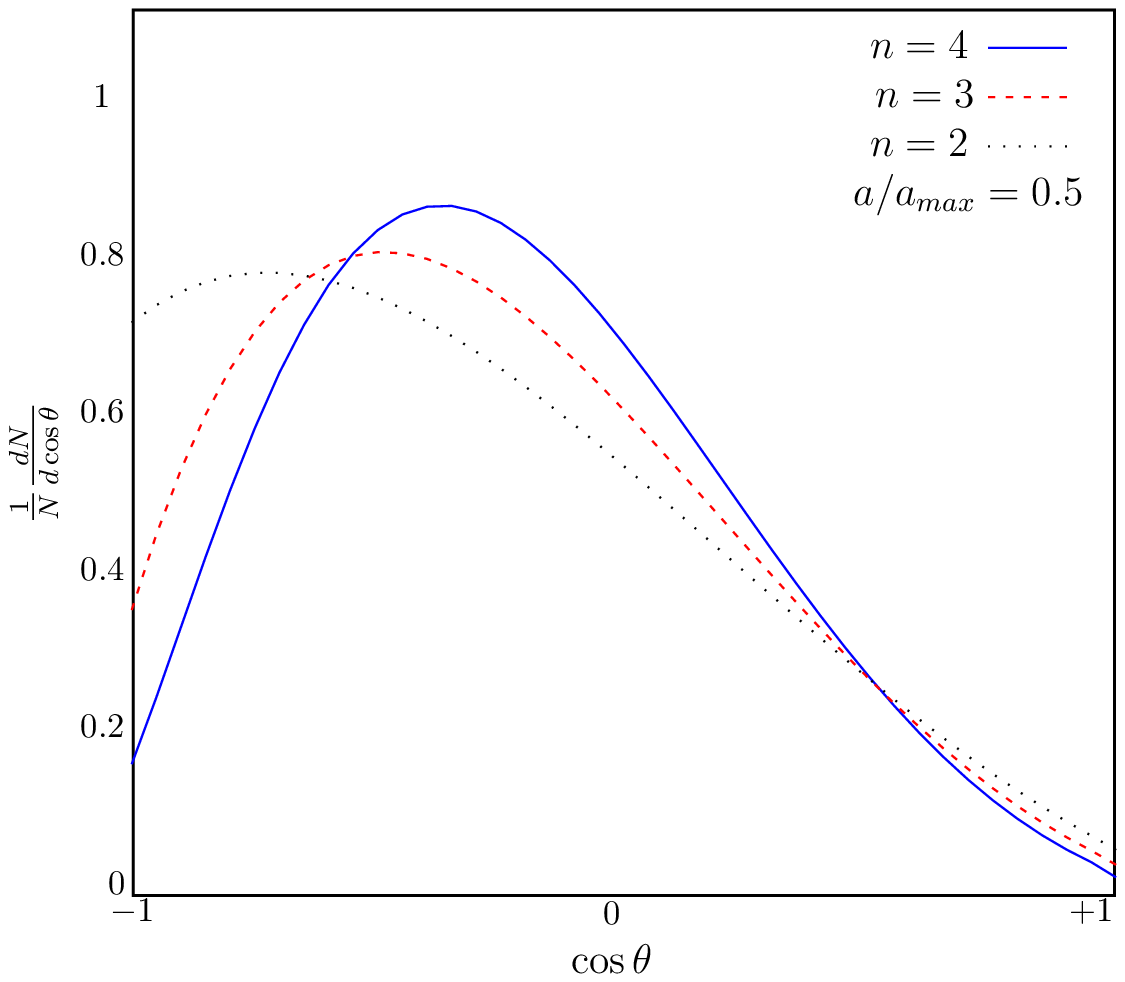}}}
\hspace{0.1cm}
\scalebox{1}{\resizebox{4.65cm}{4.80cm}{\includegraphics{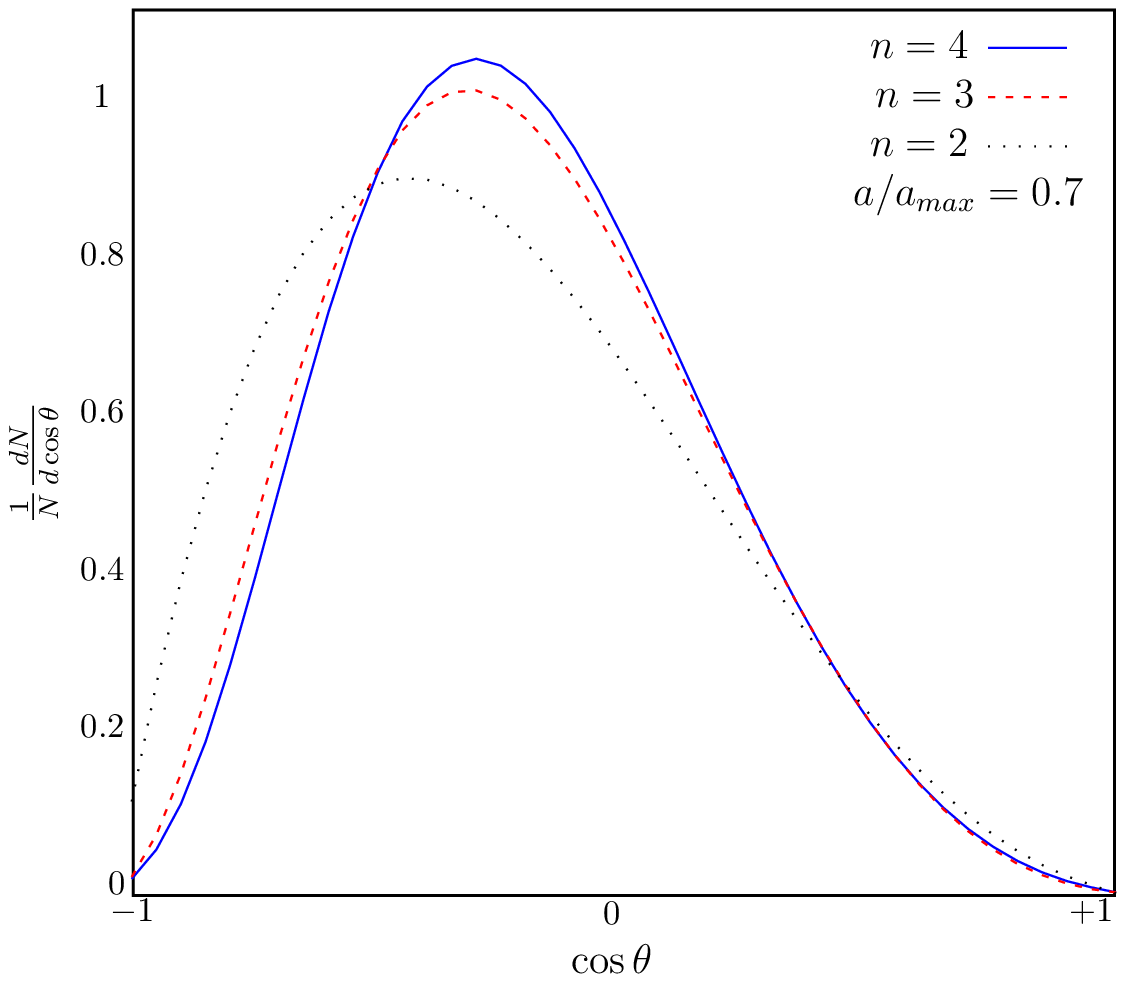}}}
\caption{
The angular distribution of emitted negative helicity fermions. 
In each plot the cases with $n=2, 3$ and $4$ are shown simultaneously. 
The reference values of the rotation
 parameter are $a_{crit}$ and $a_{max}$ in the upper and lower panels, respectively. 
The rotation parameter $a$ is set (left to right) to $30\%,~50\%,~
 70\%$ of the reference value.} 
\label{hdad}
\end{figure*}

In Fig.~\ref{hdad}, 
the angular distribution of negative helicity 
particles is displayed for various parameters, setting $\omega$ to a 
representative frequency $\bar{\omega}$. 
The value $\bar{\omega}$ is 
chosen by requiring that the fraction of particles emitted with frequency below $\bar{\omega}$, 
$N(\bar{\omega})=\int_0^{\bar{\omega}} dN$, to be 0.5.

The emission turns out to be suppressed in the direction
anti-parallel to the black hole angular momentum.
For rapid rotation, the particles tend to be emitted towards
the equatorial plane. 
This concentration in the rapidly rotating case can also 
be seen in the helicity independent angular distribution~\cite{kerr_emission_6}.  
The emission around both poles looks suppressed, 
but the observed apparent suppression is simply 
caused by the large enhancement of emission in the 
directions close to the equatorial plane. 
The asymmetry in the 
helicity dependent angular distribution is visible 
even for relatively slow rotation and becomes 
evident as $a$ increases. 
Note that, for very fast rotation, the
concentration of the emitted particles toward the equatorial plane,
observed in the angular distribution, may affect the features of cosmic
ray air showers mediated by black holes. 

For slow (rapid) rotation, 
the asymmetry decreases (increases) as the number
of extra-dimensions $n$ grows. 
This tendency may be used as an indicator to discriminate 
scenarios with different number of extra-dimensions.  
It is also worth mentioning that 
for $a/a_{crit}$ fixed
the peak position of the helicity dependent angular distribution is 
almost independent of $n$ as shown in upper panel, Fig.~\ref{hdad}. 

\vspace{1mm}
{\it Statistics.}
If we can align the direction of the axis of rotation of the black hole for
various events even approximately, we can collectively use the
experimental data to achieve high statistics for the angular
distribution of emitted particles. 
The LHC may allow to perform such measurements. 
\begin{center}
\begin{table}
\begin{tabular}{ | c  | c | c | c ||  c | c | c | c |}
  \hline
  $a/a_{max}$ &  0.3 & 0.5 & 0.7 &$a/a_{crit}$ &  0.3 & 0.5 & 0.7 \\
  \hline 
  $n=2$  &  18.20  & 15.17  & 9.47 & $n=2$  &  20.68 & 16.20 & 13.17\\
  \hline
  $n=3$ & 19.93  & 13.43  & 8.19 &  $n=3$ & 25.47  & 19.32  & 15.34\\
  \hline
  $n=4$  & 20.03  & 10.97  & 7.50  &   $n=4$  & 29.84  & 21.75  & 17.14\\
  \hline
\end{tabular}
\caption{Estimate of $\delta$ in degrees for the curves of fig.~\ref{hdad}.}
\label{table1}
\end{table}
\end{center}

\vspace{-0.9cm}
In the following, we will provide an estimation of the error in the determination of
the axis of rotation, assuming that $N$ particles are emitted per
black hole.
Let $P(\Omega)$ be the angular distribution of the emitted
particles. We expand it in terms of Legendre polynomials as 
$P(\Omega)=\sum C_{l} P_{l}\left(\cos\theta\right)$,
and consider the identification of the direction of angular momentum 
based on $l=1$ (dipole) and $l=2$ (quadrupole) moments.
For the dipole and quadrupole estimators, respectively, the errors in
the estimated direction, $\delta_d$ and $\delta_q$, can be evaluated as 
$\delta_d^2 = \frac{(1-\zeta)}{N C_{1}^2}$ and 
$\delta_q^2 = \frac{4\zeta(1-\zeta)}{N(3\zeta-1)}$, with 
$\zeta= C_0/3+2C_2/15$. Combining the dipole and quadrupole estimators, 
the $1\sigma$-error in total can be reduced to 
$\delta=\left(\delta_d^{-2}+\delta_q^{-2}\right)^{-1/2}$. 
 Assuming that the angular distribution 
shown in Fig.~\ref{hdad},
the error $\delta$ in degrees for $N=100$ is summarized in Table~\ref{table1}. 
Here we restricted our consideration to the dipole and quadrupole 
estimators, but more sophisticated statistical analyses
may reduce the error.

\vspace{1mm}
{\it Discussions.} 
In the collision of two particles at transplankian
energy, a rotating black hole is expected to form and decay. 
We studied possible
signatures of rotation of such black holes in 
departure of the energy spectra from the thermal profile, and 
in the features of the helicity dependent angular distributions. 

Black hole thermodynamics seems to suggest
the presence of an instability for $a\geq a_{crit}$. This critical 
value $a_{crit}$ is smaller
than the maximal value $a_{max}$ allowed by the kinematical requirement of
formation of a black hole in the collision of two particles. 
As far as $a\leq a_{crit}$, 
the shape of the energy spectrum 
is almost independent of $n$. 
The largest dependence on $n$ will appear in the 
effective temperature as shown in Figs.~\ref{empiricalfit}.  
However, 
this $n$-dependence must be interpreted with caution. 
When the ratio $a/a_{crit}$ is fixed, the enhancement of the effective 
temperature is larger for a smaller number of extra-dimensions. 
While the tendency is completely opposite if the ratio $a/a_{max}$ is
fixed. Hence, under the situation in which 
the true maximum value of $a$ is unknown, 
it is rather difficult to extract the information about the 
number of extra-dimensions without changing the energy of colliding 
particles~\cite{dl}. 

The peak position of the helicity dependent 
angular distribution may give a valuable information, because 
it seems to be a good indicator of $a/a_{crit}$ (or $a$ itself 
since $a_{crit}$ is always close to 1). 
Moreover, the amplitude of the anisotropy depends on the number 
of extra-dimensions. 
Hence, measuring the helicity dependent angular
distribution may provide a very important signature to extract 
the value of $n$. 
To develop analysis of this kind based on experiment, 
we need to coherently accumulate data from many events. 
For this purpose, 
it is necessary to identify the rotation axis of the formed black hole 
for each event. We have demonstrated that this identification is 
marginally possible if we can detect $O(100)$ particles. 

\vspace{-7mm}
\acknowledgments
\vspace{-5mm}
We acknowledge the support of the JSPS through Grants Nos. 19GS0219, 20740133, 17340075,
19540285 and 18204024, and Grant-in-Aid for the Global COE Program ``The Next
Generation of Physics, Spun from Universality and Emergence" from the
Ministry of Education, Culture, Sports, Science and Technology (MEXT) of
Japan. We thank Y. Sendouda for help with the numerics.

\end{document}